\documentclass[conference]{IEEEtran}
\IEEEoverridecommandlockouts
\usepackage{cite}
\usepackage{amsmath,amssymb,amsfonts}
\usepackage{algorithmic}
\usepackage{graphicx}
\usepackage{tikz}
\usepackage{hyperref}
\usepackage{lipsum}
\usepackage{caption}
\usepackage{siunitx}

\usepackage{subfig}

\usepackage{textcomp}
\usepackage{xcolor}
\def\BibTeX{{\rm B\kern-.05em{\sc i\kern-.025em b}\kern-.08em
    T\kern-.1667em\lower.7ex\hbox{E}\kern-.125emX}}

\newcommand\copyrighttext{%
  \footnotesize \textcopyright 2020 IEEE. Personal use of this material is permitted. Permission from IEEE must be obtained for all other uses, in any current or future media, including reprinting/republishing this material for advertising or promotional purposes, creating new collective works, for resale or redistribution to servers or lists, or reuse of any copyrighted component of this work in other works.}
\newcommand\copyrightnotice{%
\begin{tikzpicture}[remember picture,overlay]
\node[anchor=south,yshift=10pt] at (current page.south) {\fbox{\parbox{\dimexpr\textwidth-\fboxsep-\fboxrule\relax}{\copyrighttext}}};
\end{tikzpicture}%
}

\begin{document}

\title{End-to-End Rate-Distortion Optimization for Bi-Directional Learned Video Compression}

\author{M. Akin Yilmaz and A. Murat Tekalp \\
Department of Electrical and Electronics Engineering, Koc University,
Istanbul, Turkey \\
mustafaakinyilmaz@ku.edu.tr, mtekalp@ku.edu.tr
\thanks{This work was supported by TUBITAK project 217E033. A. Murat Tekalp also acknowledges support from Turkish Academy of Sciences (TUBA).}
}
\maketitle

\copyrightnotice

\begin{abstract}
Conventional video compression methods employ a linear transform and block motion model, and the steps of motion estimation, mode and quantization parameter selection, and entropy coding are optimized individually due to combinatorial nature of the end-to-end optimization problem. Learned video compression allows end-to-end rate-distortion optimized training of all nonlinear modules, quantization parameter and entropy model simultaneously. While previous work on learned video compression considered training a sequential video codec based on end-to-end optimization of cost averaged over pairs of successive frames, it is well-known in conventional video compression that hierarchical, bi-directional coding outperforms sequential compression. In this paper, we propose for the first time end-to-end optimization of a hierarchical, bi-directional motion compensated learned codec by accumulating cost function over fixed-size groups of pictures (GOP). Experimental results show that the rate-distortion performance of our proposed learned bi-directional {\it GOP coder} outperforms the state-of-the-art end-to-end optimized learned sequential compression as expected.
\end{abstract}
\vspace{2pt}

\section{Introduction}
\label{intro}
Conventional video compression methods, such as H264~\cite{h264} and H265~\cite{h265}, rely on human-engineered intra and inter prediction modes to exploit spatial and temporal redundancy. Inter prediction is accomplished by motion compensation based on variable size block matching. After motion compensation, residual frame is compressed using a fixed linear transformation, quantization and entropy coding steps. Rate-distortion optimization is achieved by a search strategy to select the best modes and quantization parameter based on an empirical relation between quantization step size and~rate.
Clearly, overall success of video compression depends on the quality of motion modelling, residual compression, and rate-distortion optimization. However, in the conventional framework, prediction modes and transform are predetermined, and it is not possible to optimize all free parameters simultaneously due to NP-complete nature of the optimization problem.

Inspired by the recent success of deep learning, several researchers proposed to replace motion model, transform and entropy modelling by multiple deep networks and proposed end-to-end training methods for combined optimization. The~proposed approaches include end-to-end training of sequential motion compensation~\cite{dvc} and bi-directional motion compensated prediction without end-to-end optimization~\cite{video_compress_interp}. These approaches are reviewed in more detail in Section~\ref{related}.

In this paper, we propose group-of-pictures (GoP) level end-to-end training of bi-directional motion compensated prediction for the first time and show that we achieve superior results over the prior art on learned video compression. The proposed method is presented in Section~\ref{our_work} and experimental results are shown in Section~\ref{experiments}. Section~\ref{conclusions} concludes the paper.


\section{Related work and Novelty}
\label{related}
This section relates our proposed approach to the existing literature on learned image and video compression.

\subsubsection{\textbf{Image Compression}}
Recently, end-to-end optimized learned image compression methods~\cite{balle_end}~\cite{balle_scale}~\cite{variational_low}~\cite{minnen_joint} started to achieve state-of-the-art performance superior to those of conventional image codecs, such as JPEG~\cite{jpeg}, JPEG2000~\cite{jpeg2000} and comparable to that of HEIC (intra mode of HEVC). The success of these methods can be attributed to the power of nonlinear activations and end-to-end optimization. In ~\cite{toderici}~\cite{toderici2015variable}, recurrent neural networks are used for progressive image compression. Other works employ a convolutional auto-encoder architecture~\cite{compressive_ae}. Auxiliary deep networks for rate estimation in entropy coding are proposed and generalized divisive normalization activation (GDN)~\cite{gdn} is used for better performance in~\cite{balle_end}~\cite{balle_scale}~\cite{variational_low}~\cite{minnen_joint}. In this paper, we employed the~approach of~\cite{balle_scale} for intra (key) frame coding due to its simplicity; however, we used a Laplacian model instead of a Gaussian since it yields better results.

\subsubsection{\textbf{Motion Estimation}}
In the last few years, neural network based optical flow estimation methods~\cite{flownet}~\cite{flownet2_0}~\cite{spynet}~\cite{pwcnet} started to yield state-of-the-art flow estimation on various datasets. This inspired researchers to use learned sub-pixel optical flow, instead of block-based motion vectors, in end-to-end trainable and jointly optimized video compression. In this paper, we used~\cite{spynet} for bi-directional motion estimation.

\subsubsection{\textbf{Video Compression}}
Compared to what has been achieved in end-to-end image compression, work on end-to-end optimized video compression is relatively few. We do not refer to work on improving the performance of conventional codecs using deep learning, since they are still limited by the constraints of block motion model and linear transform; hence they cannot be considered as truly end-to-end optimized.

In~\cite{video_compress_interp}, the authors propose a deep network for bi-directional compression of group-of-pictures. They follow the model used in~\cite{toderici} for compressing key (I) frames. However, for motion estimation and compensation, they provide results with both optical flow estimation by Farneb{\"a}ck's algorithm~\cite{farneback} and block motion estimation, but they do not consider end-to-end optimization of optical flow or motion vectors. 

DVC~\cite{dvc} is the first end-to-end deep video compression model that jointly learns all components of the video compression framework. Image compression, motion estimation, motion compression and residual compression components of the DVC model are all trained based on a common rate-distortion loss function. This approach gives very promising results. However, the DVC model is designed for sequential video coding, which only uses uni-directional flow vectors and the loss function is computed over pairs of frames.

Considering the fact that in conventional video compression hierarchical bi-directional motion compensation significantly improves coding performance, we propose a learned hierarchical video compression framework. Our contributions are:
\begin{itemize}
  \item We formulated hierarchical bi-directional flow estimation, flow compression, and frame prediction within a learned video compression framework.
  \item All components of the framework are designed by using learnable convolution filters and optimized in an end-to-end fashion using a single rate-distortion loss.
\end{itemize}
Experimental results demonstrate that the performance of our learned hierarchical bi-directional video compression framework is superior to that of state of the art learned sequential compression as expected~\cite{dvc}. 


\section{Proposed Method}
\label{our_work}  

\subsection{End-to-End Hierarchical Video Compression Framework}
\subsubsection{\textbf{Overview of the Model}} 
\label{overview}
\begin{figure}[b]
\centering
	\includegraphics[scale=0.270]{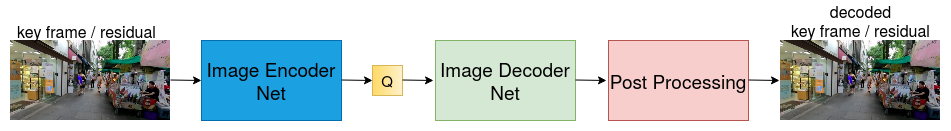} \vspace{-8pt} \\
	(a) \vspace{5pt}\\
	\includegraphics[scale=0.250]{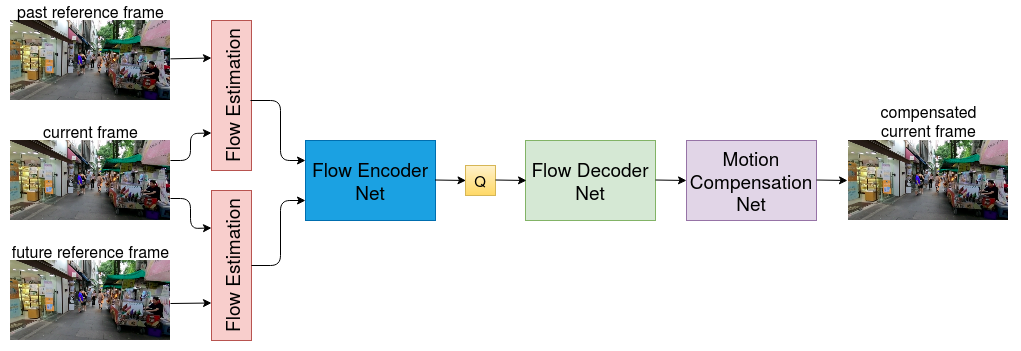} \vspace{-12pt} \\
	(b)
\caption{Block diagram of: a) intra (image) compression, where Q denotes quantization, (b) motion compensated prediction process. Note that the residual between the current frame and predicted frame is compressed using the same intra compression network in (a) re-trained separately.}
\label{fig:our_model}
\end{figure}

The proposed end-to-end learned video compression framework mimics the architecture of traditional hierarchical video compression frameworks, such as H264~\cite{h264} and H265~\cite{h265}, replacing fixed functions with trainable deep networks. We assume the GOP size N is equal to a power of 2 as usual. We briefly explain each of the sub-networks shown in Fig.~\ref{fig:our_model} below.

\subsubsection{\textbf{Image Compression Net}}
\label{imcomp}  
\begin{figure}[t]
\centering
	\includegraphics[scale=0.360]{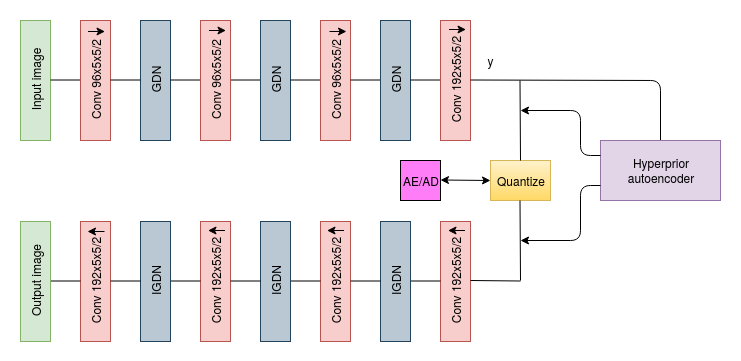} \vspace{-14pt} \\
	(a) \vspace{10pt}\\
	\includegraphics[scale=0.390]{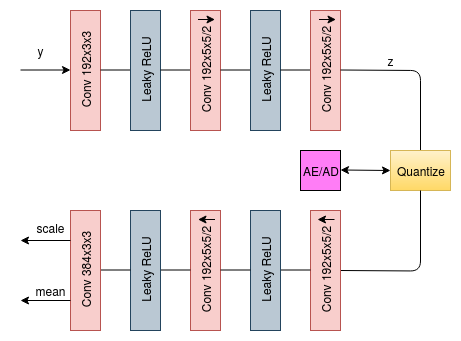} \vspace{-2pt} \\
	(b)
\caption{Block diagram of: a) image compression autoencoder, (b) the hyperprior autoencoder shown in (a). AE and AD represent arithmetic encoder and arithmetic decoder, respectively.}
\label{fig:imgcompress}
\end{figure}

We compress key frames using a CNN encoder-decoder network. This is analogous to independent compression of I frames. Key frames are compressed independently using the architecture in Fig.~\ref{fig:imgcompress}. We utilize the network based on the work~\cite{balle_scale}. The encoder and decoder are composed of multiple convolutions and GDN/IGDN nonlinearities~\cite{gdn}. The number of filters are set to 96 except for the last convolution layer, which has 192 channels. 

\subsubsection{\textbf{Flow Estimation and Compression Net}}
\label{flowcomp} After key frames are compressed and decoded by image compression network, we estimate bi-directional flow from decoded key frames to the middle frame. For flow estimation, we used \textit{Spy-Net} architecture proposed in~\cite{spynet}.  Our end-to-end compression framework, estimates flow vectors from the past and future reference frames to the current frame. Then, motion vectors are compressed using the same architecture in~\ref{imcomp}. Here, the input and the output are not images having 3 channels with shape $H \times W \times 3$, but 4 channel flow vectors (2 channels for forward and 2 for backward flow) with shape $H \times W \times 4$. 

\subsubsection{\textbf{Motion Compensation Net}}
\label{motioncmp}
Motion compensation is achieved by warping the past and future reference frames towards the current frame using the compressed and decompressed flow vectors. 

\begin{figure}[t]
\centering
	\includegraphics[scale=0.37]{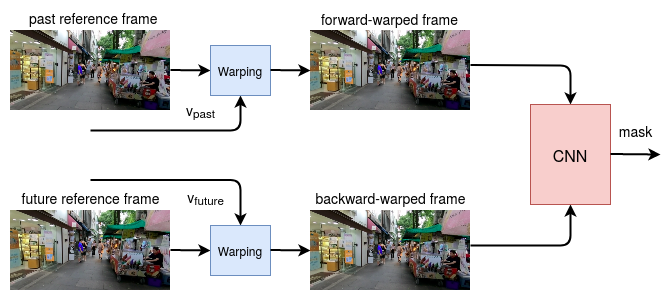} 
\caption{Block diagram of motion compensation process, which uses a CNN to estimate a bi-directional interpolation mask.}
\label{fig:fig_mc}
\end{figure}

The process, whose block diagram is shown Fig.~\ref{fig:our_model} uses a \textit{Motion Compensation Net} that finds the weights to fuse the results of forward and backward warping in order to minimize warping artifacts.
The network has a U-shaped architecture as in~\cite{voxel} to estimate sub-pixel masking for forward  and backward warping. As shown in Fig.~\ref{fig:fig_mc}, our motion compensation network takes warped future and past reference frames, then outputs a mask for each pixel. Sigmoid non-linearity is applied at the output layer of motion compensation network to have mask values between 0 and 1. Finally, refined motion compensated current frame is calculated using the mask and warped frames by
\begin{equation}
\hat{x}_{mc} = mask \times x_{pw} + (1-mask) \times x_{fw}
\end{equation}
where $\hat{x}_{mc}$ is motion compensated current frame, $x_{pw}$ is the warped past reference frame and $x_{fw}$ is the warped future reference frame. 

\subsubsection{\textbf{Residual Compression Net}}
\label{rescomp} The residual between motion compensated frame and ground truth frame is compressed using the same network architecture in~\ref{imcomp} re-trained separately. Decoded residual is then added to the motion compensated frame at the decoder side. The same procedure is applied to all frames between key frames as in~\ref{flowcomp}.   

\subsubsection{\textbf{Post Processing Net}}
\label{postprocess} 
\begin{figure}[t]
\centering
	\includegraphics[scale=0.4]{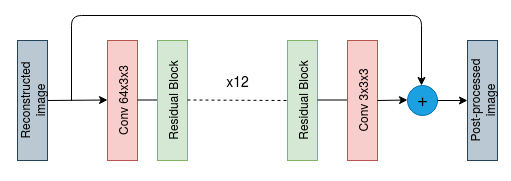} \vspace{-2pt} \\
	(a) \vspace{10pt}\\
	\includegraphics[scale=0.4]{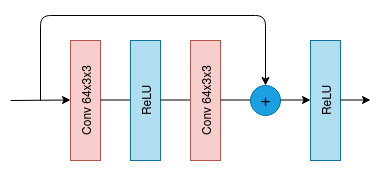} \vspace{-2pt} \\
	(b)
\caption{Block diagram of: a) post processing unit, (b) expansion of each residual block in (a).}
\label{fig:post}
\end{figure}

To reduce compression artifacts in the reconstructed images for better visual quality, a post processing unit is proposed by the authors in~\cite{variational_low}. We use 12 residual blocks with 64 channels to process reconstructed image, then add a residual connection from the image to output layer as shown in Fig.~\ref{fig:post}. We apply post processing to all reconstructed frames. 

\subsection{Training Strategy}
\label{training}

\subsubsection{\textbf{Loss Function}}
\label{loss} Our compression framework is designed to minimize the number of bits to represent a GoP while preserving the quality of decoded frames. To that effect, we use a rate-distortion loss defined as the following.\vspace{-2pt}
\begin{equation}
L = \lambda D + R_{image} +  R_{flow} + R_{residual}
\end{equation}
where $D$ denotes the distortion between original and reconstructed GoP frames. Mean square error (MSE) is used in our work for distortion loss. $R_{image}$, $R_{flow}$ and $R_{residual}$ stand for rate estimations of image, flow and residual latents respectively. $\lambda$ term in the loss function is the Lagrange multiplier that determines rate-distortion trade-off. 

\subsubsection{\textbf{Rate Estimation}}
\label{rate_estimation} Minimizing both distortion and required bits, we need to estimate rate of latent space representations. Since there is no prior information about latent space distributions, authors in~\cite{balle_scale} model each element in image latents (y in Fig.~\ref{fig:imgcompress}) with a Gaussian distribution. In their work, rate of hyper-prior latents (z in Fig.~\ref{fig:imgcompress}) is estimated by a univariate non-parametric density model. In the study~\cite{variational_low}, it is reported that modelling image latents with Laplacian distribution gives better compression performance than Gaussian modelling. Based on this evidence, we use Laplace distribution with learned hyper-priors to model image latents. Unlike previous studies, we model hyper-prior latents with unit Laplacian instead of using a non-parametric density model since we find they give very close results. 

\subsubsection{\textbf{Quantization}}
\label{quantization} Entropy coding converts latent space representations of image, flow and residual to bit stream but it accepts quantized symbols. Neural network operations must be differentiable for gradient calculations for a trainable network. Since quantization operation is not differentiable, some methods are proposed for quantization in training stage. Soft quantization approach is introduced in~\cite{soft_quantize} for differentiable quantization. Also, authors proposed to replace quantization operation by adding uniform noise during training in~\cite{balle_end}. We use uniform noise addition approach in our work. Therefore, for training, quantization of any latent space variable $\hat{y}_{i}$ is approximated as: $\hat{y}_{i} = y_{i} + \epsilon$ where $\epsilon$ is uniform noise between $-0.5$ and $0.5$. In inference stage, hard quantization is applied, i.e., $\hat{y}_{i} = round(y_{i})$.


\section{Experiments}\label{experiments}
\subsection{Experimental Setup}
\subsubsection{\textbf{Data Preparation and Evaluation}}
We train and test our compression framework using REDS dataset~\cite{reds} which is built for NTIRE 2019 video restoration challenge. It has realistic $1280 \times 720$ videos containing complex motions. We downsample that resolution to $640 \times 360$. Since our network requires each dimension to be divisible by 64, we crop all video frames to $640 \times 320$. We use group of $n = 4$ pictures for training/inference and test our model on 30 videos of REDS validation set. 
We evaluate our performance in terms of the compression rate in bits per pixel (BPP) and quality of compression in peak signal-to-noise ratio (PSNR).

\begin{figure}[t]
\centering
	\includegraphics[scale=0.60]{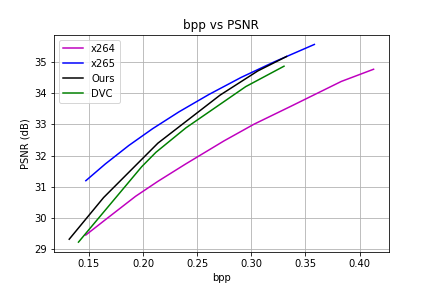} \vspace{-12pt} 
\caption{Rate-distortion (RD) comparison of our method with x264~\cite{x264}, x265~\cite{x265} and learning based codec DVC~\cite{dvc}.}
\label{fig:rd}
\vspace{-10pt}
\end{figure}

\begin{figure}[b]
\makebox[\textwidth]{\includegraphics[width=0.85\paperwidth]{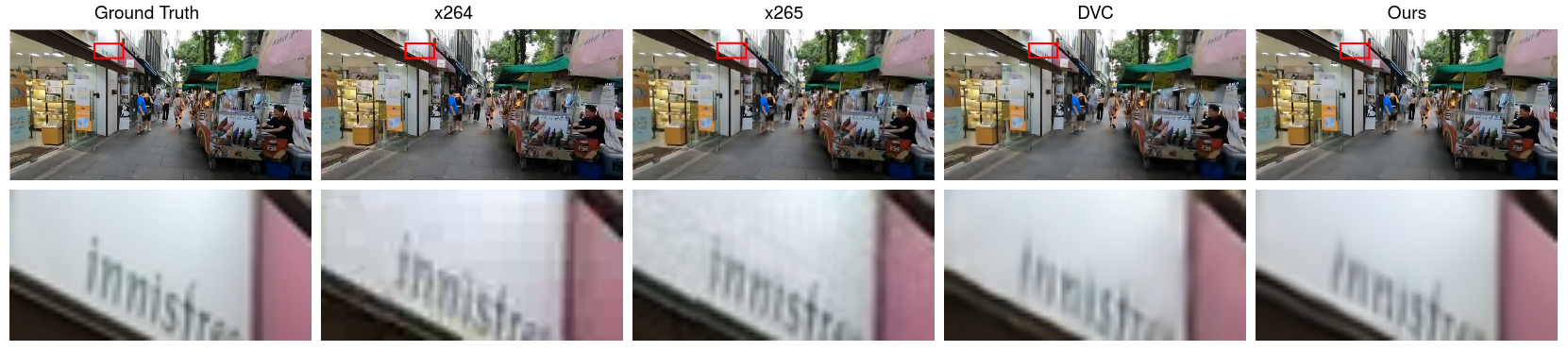}}
\caption{Visual comparison of compression results at 0.14 bpp.}
\label{fig:visual}
\end{figure}

\subsubsection{\textbf{Training Details}}
Before group of pictures level rate-distortion optimization of the whole network, we first pre-train its sub-modules. Our image compression network and post processing unit are trained on $256 \times 256$ random crops for 400K iterations using ADAM optimizer~\cite{adam} with a constant learning rate 3e-5. We apply random rotation for data augmentation and we set batch size to 16. We use pre-trained \textit{Spy-Net}~\cite{spynet} and freeze its weights to train bi-directional flow compression network. In addition to random cropping and rotation, we also use temporal flipping to augment the training data. At the last stage of training, we combine all the pre-trained modules to perform an end-to-end optimization. We train the whole framework for different \textit{$\lambda$} values in \ref{loss} to generate rate-distortion curve. The combined model is trained on randomly selected group-of-pictures with ADAM using a constant learning rate 1e-5 and batch size 4.

\subsection{Experimental Results}
Our method is compared with traditional codecs and a learning based video compression system DVC~\cite{dvc}. We use FFmpeg with \textit{ultra fast} mode to compress the frames using commercial x264~\cite{x264} and x265~\cite{x265} codecs. The GOP size is set to 4 for a fair comparison. Fig.~\ref{fig:rd} shows the experimental results on REDS validation set. In the low bitrate regions, far behind x265, our method gives similar results to x264. As moving towards higher bitrate, the difference between our system and x264 is widened, and eventually our curve intersects with that of x265. Across the entire bitrate range, our method outperforms the recent learning based work~\cite{dvc} by about 0.2 to 0.5 dB. It is a clear evidence that, utilization of bi-directional frame prediction improves the coding performance compared to the uni-directional prediction.

Fig.~\ref{fig:visual} reports visual quality comparison of different compression algorithms at nearly the same birate, 0.14 bpp. For different compression systems, the PSNR of reconstructed frames is measured as x264: 29.30 dB, x265: 31.03 dB, DVC: 29.25 dB, Ours: 29.85 dB. Without any blocking artifacts, our method offers a smoother visual result compared to x264 and x265 despite it achieves a lower PSNR than x265. Our method gives the second highest PNSR (29.86 dB) behind x265 (31.03 dB). Both traditional codecs bring some blocking artifacts in the texture when zooming in to the cropped sections. On the other hand, learning based methods show smoother results while not having the same blocking artifacts. 


\section{Conclusions and Future Work}
\label{conclusions}
We propose a promising end-to-end optimized hierarchical bi-directional learned deep video compression framework, where we minimize the average of the rate-distortion cost function accumulated over groups of pictures.
Our results exceed the performance of hierarchical bi-directional coding using x264 and end-to-end optimized learned sequential coding~\cite{dvc} and comes very close to those of x265 codec at higher bitrates.

Since the proposed approach is highly modular, new results in learning based optical flow estimation and context modeling in entropy coding can easily be plugged into our framework for improved performance in the future. For example, a context model for exploiting the distribution of the latents as proposed in~\cite{minnen_joint} can further improve compression performance. 


\clearpage
\bibliography{references}
\bibliographystyle{IEEEtran}
\end{document}